# Rigorous numerical study of strong microwave photon-magnon coupling in all-dielectric magnetic multilayers

Ivan S. Maksymov[a)], Jessica Hutomo, Donghee Nam and Mikhail Kostylev

*School of Physics M013, The University of Western Australia, 35 Stirling Highway, Crawley WA 6009, Australia*

[a)]ivan.maksymov@uwa.edu.au

**Abstract**: We demonstrate theoretically a strong local enhancement of the intensity of the in-plane microwave magnetic field in multilayered structures made from a magneto-insulating yttrium iron garnet (YIG) layer sandwiched between two non-magnetic layers with a high dielectric constant matching that of YIG. The enhancement is predicted for the excitation regime when the microwave magnetic field is induced inside the multilayer by the transducer of a stripline Broadband Ferromagnetic Resonance (BFMR) setup. By means of a rigorous numerical solution of the Landau-Lifshitz-Gilbert equation consistently with the Maxwell's equations, we investigate the magnetisation dynamics in the multilayer. We reveal a strong photon-magnon coupling, which manifests itself as anti-crossing of the ferromagnetic resonance (FMR) magnon mode supported by the YIG layer and the electromagnetic resonance mode supported by the whole multilayered structure. The frequency of the magnon mode depends on the external static magnetic field, which in our case is applied tangentially to the multilayer in the direction perpendicular to the microwave magnetic field induced by the stripline of the BFMR setup. The frequency of the electromagnetic mode is independent of the static magnetic field. Consequently, the predicted photon-magnon coupling is sensitive to the applied magnetic field and thus can be used in magnetically tuneable metamaterials based on simultaneously negative permittivity and permeability achievable thanks to the YIG layer. We also suggest that the predicted photon-magnon coupling may find applications in microwave quantum information systems.

## I. INTRODUCTION

Unique electromagnetic properties of microwave waveguides and resonators with magneto-insulating material inclusions have been well-known and actively exploited for several decades [1]. Recently, the functionality of devices comprising magneto-insulating materials has been extended and now it also includes spin wave-based and magnonic microwave devices [2–4], magnetically tunable microwave metamaterials [5–22] and microwave quantum systems [**23–28**]. It is worth stressing the practical importance of microwave quantum systems. Although in some cases the realisation of quantum functionality in the microwave range is more complicated than in optics, microwave quantum systems are potentially more useful in the short-term perspective because they open up opportunities to increase the sensitivity of magnetic-resonance imaging and improve the currently available radar technologies (see, e.g., Refs. [**29, 30**]).

Both the tuneability effect in metamaterials and the strong coupling of microwave photons to magnons (the quanta of spin waves [**2–4**]) in quantum systems stem from the dependence of the ferromagnetic resonance (FMR) frequency on the external static field $H$ applied to the magnetic material [**1**]. This coupling leads to a strong anti-crossing between the microwave photon mode and the magnon mode. Recently, Cao et. al. [**26**] demonstrated that the anti-crossing can be realised not only for the FMR, but also for higher-order standing spin wave resonances co-existing with the FMR in a magnetic thin film. As these resonances exist at lower value of the applied magnetic field, this finding opens up opportunities to decrease the applied field required for the strong coupling.

A split-ring resonator (SRR) is arguably one of the most often employed elements in metamaterials in general (see, e.g., Ref. [**31**]) and in magnetically tunable microwave metamaterials in particular (see, e.g., Refs. [**14–19**]). SRR-based devices represent a magnetic film-loaded SRR that is excited by a microstrip or coplanar line (for brevity and generality, hereafter we also use the term "stripline"). In the case of the excitation of the SRR by a stripline [**32**], the dynamic magnetic field $h_{rf}$ created by the stripline on top of which the SRRs loaded with the magnetic film are located, is used to drive the electromagnetic modes of the SRR and the magnon modes in the magnetic film. (Hereafter, we will interchangeably use the terms "electromagnetic mode" and "photon mode".)

On the other hand, there are alternatives to SRR-based magnetically tuneable devices. For example, Yttrium Iron Garnet (YIG) thin films or bulk YIG also exhibit negative permittivity and permeability [**5–8, 10, 21, 22**]. Such YIG-based devices are often excited by the dynamic magnetic field $h_{rf}$ of a stripline and their operating frequency can be tuned by an applied static magnetic field. YIG is a very well-known material that have been used in industry for several decades [**4, 33**]. YIG has a low magnetic (Gilbert) damping parameter $\alpha_G = 0.001$ (or even lower, see Ref. [**2**]), which is eight times smaller than that of Permalloy, an alloy having one of the smallest $\alpha_G$ among the ferromagnetic metals [**3**].

Most significantly, YIG is very attractive for microwave quantum experiments. It is known that whereas the interaction of a single spin with an electromagnetic field mode is very weak, collective enhancement, which scales as the square root of the total number of spins [**34**], makes the effective coupling strong enough. The spin systems with the largest volumetric density of spins are ferro- and ferrimagnets. For instance, the net spin density of YIG in the ordered phase is several orders of magnitude higher than in paramagnetic spin ensembles [**33**], which have been recently studied extensively and shown to be prospective candidates for quantum information applications (see, e.g., Ref. [**35**]). Furthermore, in a ferro- or ferrimagnetic materials the spins are strongly coupled to each other which results in a collective motion of spins required for the strong coupling.

In this work, we propose and investigate a multilayered structure having in its core a magneto-insulating YIG film. The YIG layer is sandwiched by non-magnetic low-loss high dielectric constant (HDC) layers. The in-plane dimensions of the HDC layers match those of the YIG film and these layers have a large dielectric permittivity matching that of YIG [36]. We also consider the scenario of a bulk YIG layer used instead of the film. The HDC layers are not required in this scenario.

By means of rigorous numerical simulations we predict a strong microwave photon-magnon interaction, which manifests itself as anti-crossing between the magnon mode and the fundamental electromagnetic resonance mode supported by the multilayer. As the strength of the anti-crossing effect depends on the applied static magnetic field, we suggest that our multilayered structure can be employed as a platform for microwave planar tunable metamaterials and quantum systems. In the latter case, our theoretical predictions are for a signal level well above the single-photon level (classical dynamics). This valid approach was pointed out, e.g., in Ref. [24], which demonstrated that the prospective candidates for high-co-operativity systems can be identified by studying their classical dynamics at room temperature.

Similar to non-tuneable all-dielectric microwave metamaterials (see, e.g., Refs. [9, 37, 38], the proposed multilayered structure are expected to be low-loss as compared with SRR-based devices, which are not free of losses due to capacitance gaps and currents flowing in the metallic parts of the SRR. Most significantly, we deliberately design our multilayered structure to resonantly enhance the microwave magnetic field at the frequency of the FMR in the YIG film. As shown in Refs. [37, 39], the local enhancement of the microwave magnetic field is due to the presence of a Mie-type resonance in subwavelength dielectric structures, i.e. structures with cross-sectional dimensions smaller than the microwave wavelength.

The enhancement of the microwave magnetic field $h_{rf}$ attainable with our structure is advantageous for magnetically tuneable microwave devices because this field drives the FMR resonance. Consequently, $h_{rf}$ has to be strong enough to achieve an easily detectable FMR response. Whereas this is not a problem to achieve a strong response in cavity FMR measurements employing a high quality factor microwave cavity, $h_{rf}$ is usually weak in stripline BFMR and thus it needs to be enhanced. For instance, $h_{rf}$ might be increased by using a narrower stripline. However, in this case the near field produced by the stripline is mainly concentrated above the stripline in close proximity to its surface. This low-lying field localisation makes it difficult to drive magnetisation dynamics in thick magnetic films. Moreover, by using a narrow stripline one also excites propagating spin waves, which in general are seen as an undesirable widening of the FMR linewidth [40, 41]. We suggest that by choosing our all-dielectric multilayered structure one can increase the amplitude of $h_{rf}$ virtually independently of the dimensions of the stripline.

All-dielectric resonators combining magneto-insulating bulk YIG or thin films YIG with non-magnetic dielectric materials were used in the past. Thus, before we proceed to the discussion of the main results of our work, we would like to additionally highlight the novelty of our results with respect to those presented in previous works:

1) Previous works Refs. [**20, 42–51**] demonstrated the existence of spin-electromagnetic waves in multilayered structures consisting of a YIG film combined with a ferroelectric or piezoelectric materials. In such structures, a spin-electromagnetic wave results from anti-crossing interaction between a spin wave in the YIG film and an electromagnetic wave propagating mostly in the electrically controlled layer. Consequently, the main discussion in the cited papers was focused on the electric control of the spectrum and phase shift of the spin-electromagnetic wave. In addition to electrical tuning, it was shown that the aforementioned multilayered structures can also be controlled by applied static magnetic fields (see, e.g., Refs. [**45, 51**]). However, none of the cited papers provided a detailed discussion of the mode anti-crossing phenomenon as a function of the applied magnetic field that is being varied in a broad range. Moreover, in some studies the mode anti-crossing effect was considered as undesirable [**48**]. Finally, the cited papers also do not discuss microwave magnetic field enhancement properties of the multilayered structures. A reason for this may be a very large (~1500) dielectric constant of ferroelectric layers as compared with the permittivity of the HDC material used in this present work. For such large permittivity, the fundamental resonance shifts toward low frequencies and thus in the spectral range of interest one does not observe electromagnetic field resonances suitable for the field enhancement.

2) In a recent work Bi et. al. [**10**] proposed an arrayed magnetically tuneable Mie-type resonance-based dielectric metamaterial, which consists of individual meta-atoms built of dielectric cube resonators surrounded by bulk YIG layers. Bai et. al. investigated the effective permeability and permittivity of the metamaterial and showed the possibility to control these parameters by external static magnetic field. However, despite a close similarity between their metamaterial structure and the multilayered structure proposed in this work, one can easily see important differences. Firstly, we propose a single (as opposed to an array) planar YIG thin film-based structure capable of enhancing the in-plane component of the microwave magnetic field $h_{rf}$. However, the microwave magnetic field enhancement was not discussed by Bai et. al. Most significantly, they do not discuss the mode anti-crossing effect that lies at the heart of magnetically tuneable materials and microwave quantum systems [**16, 17, 23–28**]. In fact, the anti-crossing could not be observed in the experimental arrangement used by Bai et. al. because at the maximum of the static magnetic field applied normally to the metamaterial (~2500 Oe) the FMR frequency was below the resonance frequencies observed in the transmission spectrum of the metamaterial.

3) A high-quality dielectric resonator was used in Ref. [**52, 53**] to create pumping field for the amplification of standing spin wave modes in a single crystal YIG film. This resonator was excited at a fixed carrier frequency of 14.258 GHz, which corresponds to the double frequency of one of the standing spin wave modes. This arrangement is required to obtain the maximum efficiency of the parametric amplification. Consequently, the anti-crossing between the spin wave modes and the dielectric resonator mode was not reported in the cited papers. Importantly, one should not mix up the aforementioned anti-crossing effects with the hybridisation of different types of spin wave modes observed in Fig. 1(b) in Ref. [**52**].

4) Finally, the papers cited in (1)–(3) do not discuss the potential of the strong photon-magnon coupling observed in all-dielectric multilayers to be applied in microwave quantum systems. In this work, we point out this potential and also suggest that the use of dielectric materials may help to decrease losses in metallic cavities and SRRs used in the previous works [**23–25, 27**]. A conceptually similar idea has recently been proposed by Rameshti et. al. [**54**] who studied a YIG sphere placed inside a spherical microwave cavity. They demonstrated that one can achieve strong photon-magnon coupling even if the cavity is removed and that the isolated YIG sphere itself acts as a microwave cavity that supports Mie-type resonances. However, YIG spheres are essentially 3D objects which are difficult to use in a planar configuration, which is not the case of all-dielectric multilayers investigated in our work.

## II. RESULTS AND DISCUSSION

Central for understanding of the FMR is the notion of spin waves, which are excitations in magnetic media existing in the microwave frequency range [**1, 2**]. Spin waves represent collective precessional motion of spins coupled by short-range exchange and long-range dipole interaction in a magnetic medium. The classical description of spin waves is given by the Landau-Lifshitz-Gilbert (LLG) equation for the magnetisation vector $\mathbf{M}$

$$\frac{\partial \mathbf{M}}{\partial t} = -|\gamma|\left(\mathbf{M} \times \mathbf{H}_{\text{eff}}\right) + \frac{\alpha_{\text{G}}}{M_{\text{s}}}\left(\mathbf{M} \times \frac{\partial \mathbf{M}}{\partial t}\right),$$

(1)

where $\gamma$ is the gyromagnetic ratio, $\mathbf{H}_{\text{eff}}$ is the total effective magnetic field inside the medium including the contribution of $h_{\text{rf}}$, $M_{\text{s}}$ is the saturation magnetisation, and $\alpha_{\text{G}}$ is the Gilbert damping coefficient. The first term on the right–hand–side of Eq. (1) gives rise to the precessional motion of the magnetisation vector about an equilibrium direction determined by the effective magnetic field. The second term is the damping term responsible for the magnetisation vector spiralling back to the static equilibrium.

The FMR or the fundamental mode of uniform precession of magnetisation is the case where the spins precess with the same phase and amplitude over the whole volume of the magnetic material. It may be considered as a spin wave with an infinite wavelength or zero wave vector.

The most popular way to probe the excitation of the FMR in planar multilayered magnetic structures is to expose a planar sample to a microwave radiation by using stripline broadband ferromagnetic resonance (BFMR) method (for a review see, e.g., Ref. [**41**]). The main part of a BFMR setup is a section of a stripline. The multilayer under test sits on top of the stripline (**Fig. 1**). A microwave current flowing through the microstrip at a fixed frequency $f$ imposes a microwave (Oersted) magnetic field on the multilayer. The resonance frequency in the multilayer is determined by a slowly scanned frequency $f$ or external static magnetic field $H$. In the latter case, as the value of $H$ is adjusted, the frequency of the natural magnetisation precession resonance eventually equals the frequency of the microwave magnetic field, and significant microwave power absorption occurs.

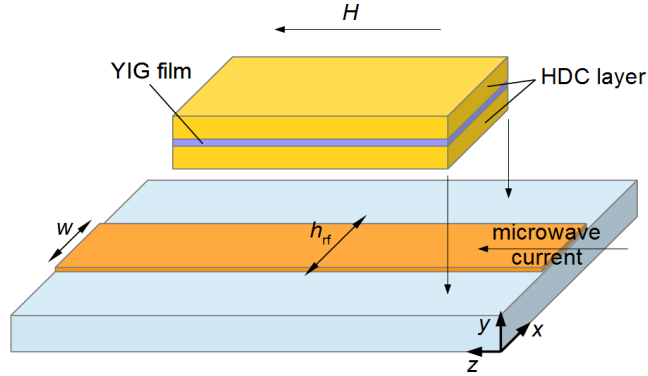

**FIG. 1** Schematic of the proposed multilayered structure consisting of a magnetic YIG layer sandwiched by two non-magnetic high dielectric constant (HDC) layers. (The GGG substrate of the YIG film is not shown.) The cross-sectional dimensions of the multilayer are 4.65 mm along the $y$-axis and 8 mm along the $x$-axis. In simulations, it is assumed that the multilayer is infinitely long along the $z$-axis and the YIG layer is magnetised tangentially by the external static magnetic field $H$. The microwave magnetic field $h_{\mathrm{rf}}$ is induced by a microwave current flowing in the stripline of width $w$. In the main text, the $x$-component of this field is referred to as in-plane. Only for the sake of illustration, the multilayered structure is separated from the stripline by an air gap. In practice, and thus in our simulations, this structure sits on top of the stripline, as indicated by two straight downward arrows.

The core of the proposed multilayered structure is a magneto-insulating 50 µm-thick YIG film (**Fig. 1**, the thickness is defined along the $y$-axis). The YIG film sits on top of a thicker (~0.5 mm) non-magnetic Gadolinium Gallium Garnet (GGG) substrate, which is used as a seed for the growth of YIG. Hereafter, we will assume that in the microwave spectral range the dielectric permittivity of both YIG and GGG is 15. We sandwich the YIG/GGG film by two non-magnetic HDC plates with the same in-plane dimensions as those of the YIG/GGG film. Importantly, we assume that the dielectric permittivity of these commercially available [**36**] low-loss HDC plates is also 15. Thus, in the cross-section our final structure looks like a 4.65 mm-high and 8 mm-wide multilayer consisting of a magneto-insulating YIG film surrounded by non-magnetic HDC materials.

As discussed above, in an experimental work the BFMR spectroscopy would be the method of choice with which to investigate the FMR response of the multilayered structure. In Ref. **41** we showed that experimental BFMR traces can be reproduced with high accuracy and explained by using semi-analytical and numerical methods such as a finite-difference method or a finite-difference time-domain (FDTD) method. In this work we employ our customised FDTD software that solves the LLG equation Eq. (1) consistently with the Maxwell's equation [55]. Due to its time-domain nature, the FDTD may be thought as a numerical counterpart of the experimental Pulse Inductive Microwave Magnetometry (PIMM) technique [41], which is one of the experimental techniques belonging to the family of the BFMR spectroscopy.

We would like to underline the complexity of numerical simulations due to the presence of a very thin magnetic YIG film as compared with the microwave wavelength of $h_{rf}$ (see Ref. [56] for more details). For this reason, hereafter we will assume that the multilayer is infinitely long along the $z$-axis. Also, we will model the microstrip line as an infinitesimally thin current sheet. This approach was used in the past in analytical models (see, e.g., Ref. [57]). We also demonstrated that the aforementioned two-dimensional model and the current sheet approximation reproduce experimental data with good accuracy (see Ref. 41 and references therein).

Another numerical problem arising due to the large difference between the film thickness and the wavelength of the electromagnetic wave [56] – the instability of Perfectly Matched Layers boundary conditions for the FDTD – has been circumvented in this work by using the old-fashioned Mur's absorbing boundaries [58], which have been proven to be acceptably efficient.

In our simulations we use the following parameters of YIG: $4\pi M_s = 1750$ G and $\alpha_G = 0.001$. Thus, due to a small saturation magnetisation and a relatively large thickness of the YIG film the higher-order standing spin waves are not expected to contribute to the magnetisation dynamics. As the dipole-dipole interaction dominates the total energy, our FDTD simulations do not take into account the exchange interaction, which makes it possible to reduce the computation time. Also, some authors [2] cite smaller values of $\alpha_G$ for YIG, which, if used in our simulations, would lead to very small values of the FMR linewidth $\Delta H$. Thus, such values are deliberately avoided in our simulations because the resulting resonance peaks would be very sharp and the attainment of high resolution of these peaks would be very time-consuming.

As a test of our software and also in order to chose a suitable discretisation parameters of the computation domain $\Delta x$ and $\Delta y$, we conduct simulations for the standalone 50 μm-thick YIG film located 2.325 mm above the microstrip line, which simulates the scenario of the surrounding material with the dielectric permittivity of air. The YIG film is magnetised in its plane along the $z$-axis (Fig. 1).

We excite the magnetisation dynamics in the YIG film by modelling the excitation of the microstrip line by microwave current. As a rule of thumb, the near field created by a microstrip line extends above this line at a distance that equals to the width $w$ of the microstrip. Consequently, we chose the width of the microstrip line as $w = 3$ mm, which ensures a strong coupling between the microwave magnetic field of the stripline and the FMR resonance in the YIG film. The choice of a such a wide stripline also ensures an efficient excitation of the electromagnetic resonance in the case of the YIG film surrounded by the HDC material (see below).

We also note that the excitation by using such a wide stripline is equivalent to the excitation by a plane wave incident on the sample from the far-field region [**41**], which was additionally verified in this work. Consequently, the result presented below may be extended to the scenario of the excitation from the far field region, which is often the case of electromagnetic metamaterials (see, e.g., Ref. [**10**]).

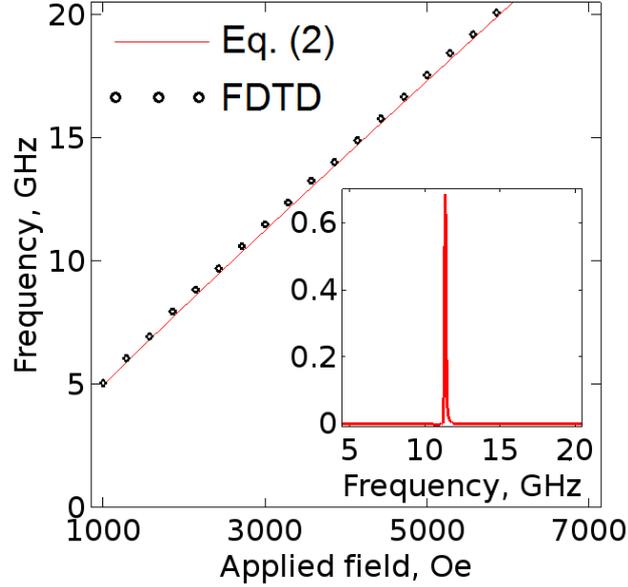

**FIG. 2** Simulated FMR response of the isolated 50 μm-thick YIG film (circles) and the prediction of the Kittel equation Eq. (2) (thin solid line). The inset shows the FMR spectrum (in arb. units) at $H$ = 3000 Oe.

As shown in **Fig. 2**, we obtain a good agreement between the simulated FMR frequency and the predictions of the Kittel equation [**1, 2**]

$$f_{\mathrm{FMR}} = \frac{|\gamma|}{2\pi} \sqrt{H\left(H + 4\pi M_{\mathrm{s}}\right)}. \tag{2}$$

We note a good agreement between the results in a broad range of frequencies and values of the applied field. This results was achieved by using $\Delta x$ = 62.5 μm and $\Delta y$ = 2.5 μm. The total number of time steps was considerably large: $10^6$. For the sake of consistency, in all simulations presented below we will keep the same parameters.

The obtained FDTD spectra also allow us to extract the frequency swept FMR linewidth $\Delta f$, which is then used to obtain the field swept full width $\Delta H$ of the FMR [**59**]. We note that the extraction of the linewidth is challenging because of sharp features in the spectra (see the inset in **Fig. 2**), which is consistent with the fact that in our model we use a relatively small Gilbert magnetic damping constant $\alpha_{\mathrm{G}}$ = 0.001. Nevertheless, we achieve acceptable agreement between the extracted linewidth values and the values for an infinitely wide YIG film produced by the formula $\Delta H = 2\alpha f/|\gamma|$ [**2, 60**].

This result allows us to assume that the resulting linewidth is not affected by the broadening due to the excitation of travelling spin waves [**40, 41**]. Indeed, the previous theory predicts that this broadening should be negligible because our stripline is wide enough to minimise this effect. Furthermore, we suggest that another linewidth broadening mechanism – the excitation of the higher-order width modes [**61**] – is also absent.

The width modes have the wave vectors $k = n\pi/w_{film}$ being $n$ the mode number indices and $w_{film} = 8$ mm the width of the magnetic film (along the $x$-direction in **Fig. 1**). As shown in Ref. [**41**], in the stripline BFRM one can excite spin waves with the maximum wave vector $k_{max} = 2\pi/w$ (where $w$ is the characteristic width of the stripline). In fact, $k_{max}$ is the first zero of the function $j_k = (w/(2\pi))\sin(kw/2)/(kw/2)$, which corresponds to the Fourier image of the linear current density in the stripline. Thus, the spectral density of $j_k$ is largely concentrated between $-k_{max}$ and $k_{max}$. For example, for $w = 3$ mm one obtains $k_{max} \approx 21$ cm$^{-1}$.

Due to symmetry considerations, in our structure one may expect to observe a noticeable contribution of the width modes with $n = 1, 3,$ and, $5$ to the FMR spectra because their wave vectors $k < k_{max}$. However, as shown in Refs. [**57, 62**], in the stripline arrangement the amplitude of the driving microwave field scales as $\exp(-kd)$ being $d$ the distance between the stripline and the film, which in our case is 2.325 mm. By plugging $d$ and $k = n\pi/w_{film}$ into this expression one sees that the efficiency of the excitation of the width modes with respect to the fundamental FMR mode quickly drops with the mode number $n$, which supports our hypothesis that the higher-order width modes are absent in our system.

As the next step, in our simulations we sandwich the YIG film between two HDC plates (as shown in **Fig. 1**). Firstly, we calculate the spectrum of the multilayer for the zero applied static magnetic field $H$. The calculated spectrum [**Fig. 3(a)**] shows the ratio $|h_x|^2/|h_0|^2$ taken in the middle of the YIG film, where $h_x$ is the in-plane component of the microwave magnetic field induced by the stripline loaded with the multilayer and $h_0$ is the in-plane field component of the unloaded stripline [**63**]. One can see a resonant >350-fold local enhancement of the intensity of the in-plane ($h_x$) component of the microwave magnetic field at the microwave frequency of 11.2 GHz. One can see that the linewidth of this peak is one order of magnitude larger as compared with the FMR linewidth. The obtained enhancement is consistent with a previous result for a single dielectric cuboid resonator with a similar dielectric constant but larger dimensions [**39**].

Figure **3(b)** shows the simulated profile of the $h_x$ field, from which one sees that the maximum of the local enhancement occurs slightly above the centre of the multilayer, i.e. slightly above the YIG film whose approximate position is given by two straight horizontal lines. This shift of the field profile is probably due to one-side excitation of the multilayer by a stripline, which is located below the multilayer. We verified that the same shift is seen when we excite the multilayer by a plane wave incident from the bottom.

It is also noteworthy that virtually the same spatial profile of the $h_x$ field (but not the amplitude) is observed when the frequency is scanned from 8 GHz to 14 GHz. We will return to this discussion later on.

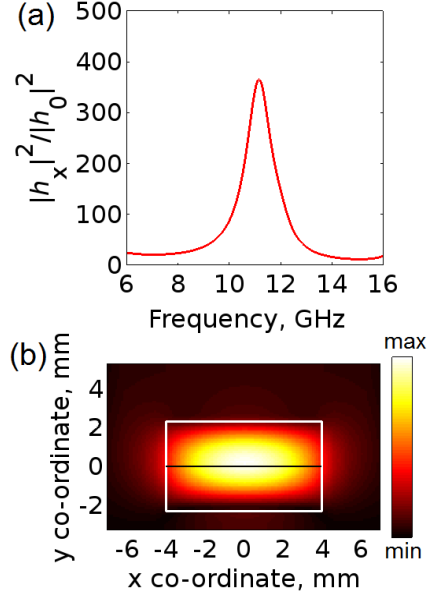

**FIG. 3** (a) Local intensity enhancement of the in-plane microwave magnetic field component in the *non-magnetised* multilayered structure as a function of the frequency. The thickness of the YIG film is 50 μm, which is approximately indicated by two parallel horizontal black lines in Panel (b). In simulations, the co-ordinate of the field detector is $x = 0$, $y = 0$. (b) Simulated $|h_x|$-field profile at the resonance frequency 11.2 GHz. The white rectangle denotes the contour of the multilayer. The stripline is located under the multilayer.

As the next step, we step up the value of $H$ and repeat the simulation of the $h_x$-field enhancement spectrum. As shown in **Fig. 4**, the peak corresponding to the magnon mode appears at ~8.2 GHz at $H = 1999$ Oe. The frequency of this peak increases as the value of $H$ is increased. At around $H = 2900$ Oe, the magnon mode peak reaches the electromagnetic mode peak and the anti-crossing occurs. In this case one can see not one but two peaks at the frequencies above and below 11.2 GHz. The magnon mode peak reappears at ~13 GHz at $H = 3571$ Oe, and the position and magnitude of the electromagnetic resonance peak of the multilayer take the same values as before the anti-crossing occurred. Figure **5(a)** provides a bird's-eye view picture of the anti-crossing effect discussed in **Fig. 4**. This figure shows the simulated spectra as a two-dimensional gray-scale map plotted as a function of both frequency and the applied field $H$.

We also extract $\Delta H$ of the FMR peak at the frequencies far from the anti-crossing by using the same procedure as above. The obtained values resemble those obtained for an infinitely wide YIG film but without the HDC layers. At this point we would like to return to the discussion of the width modes and the role of the profile of the microwave magnetic field inside the YIG film. In accord with the theory from Guslienko et. al. [**61**], the inhomogeneity of the $h_x$-field profile along the width of the magnetic film (along the *x*- direction in **Fig. 1**) is a condition for efficient excitation of the width modes. We already discussed that a considerable distance between the stripline and the YIG film drastically reduces the efficiency of the excitation of these modes in the isolated YIG film. Here, we point out that the presence of the HDC layers additionally prevents these modes from occurring. Above we noted that the $h_x$-field profile shown in **Fig. 3(b)** virtually remains unchanged in a broad range of frequencies from 8 GHz to 14 GHz. Therefore, one sees that the presence of the HDC layers makes the field profile less inhomogeneous, thus making this field unfavourable for the excitation of the width modes.

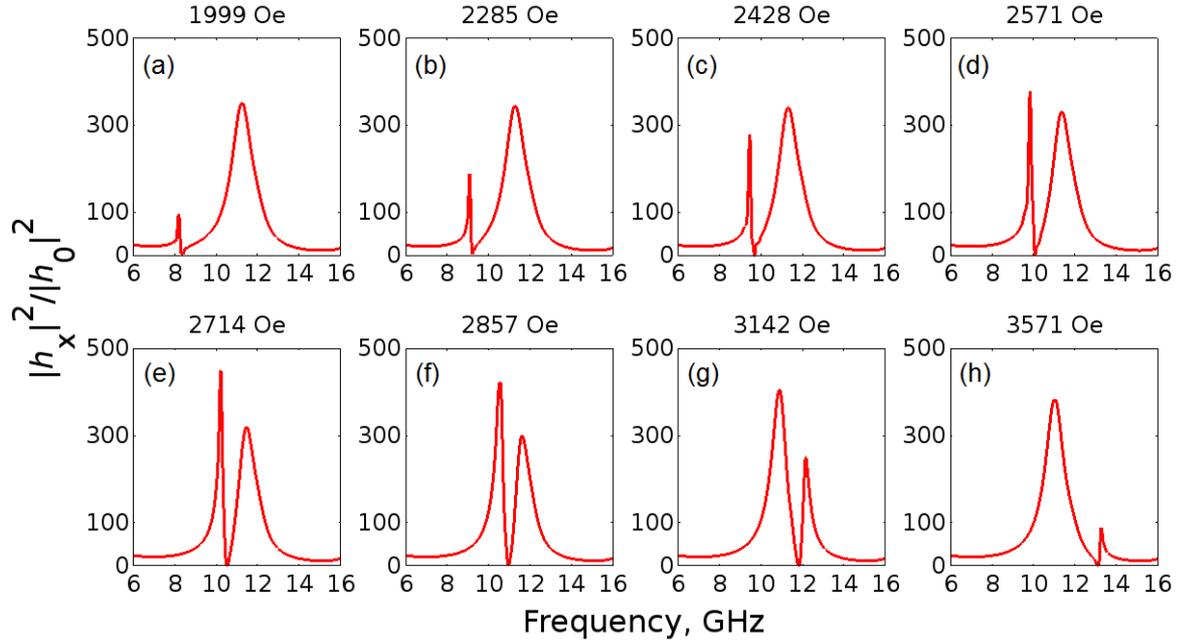

**FIG. 4** (a-e) Simulated spectra of the multilayer as a function of the frequency for different values of the applied static magnetic field. The thickness of the YIG film is 50 µm. In simulations, the co-ordinate of the field detector is $x = 0$, $y = 0$.

By analogy with SRR-based metamaterials loaded with bulk YIG [**16**], it would be logical to expect a larger magnon mode peak intensity in thicker YIG layers sandwiched between two HDC layers. On one hand, the strength of the interaction between spins and an electromagnetic field mode scales as the square root of the total number of spins [**34**]. Consequently, the number of spins is larger in a thicker YIG layers than in thinner ones, which should result in stronger interaction. On the other hand, an earlier theory from Ref. [**57**] states that the strength of coupling of magnon dynamics in ferromagnetic films to microwave fields of striplines scales with the film thickness.

However, the previous theories do not take into account the contribution of the non-magnetic HDC layers present in the multilayer. As one can imagine by inspecting **Fig. 3(b)**, a larger portion of the microwave magnetic field energy can be focused in the cross section of a thicker YIG layer as compared with a thinner one. This suggests that the strength of the FMR excitation in thicker layers can be yet stronger due to the localisation of the microwave magnetic field in the layer.

The result of numerical simulations for a 500 μm-thick YIG slab is shown in **Fig. 5(b)**. Note that we do not change the total thickness of the multilayer, which implies that for the 500 μm-thick YIG film each surrounding HDC layer is 225 μm thinner than for the structure with the 50 μm-thick YIG film. One observes a stronger anti-crossing effect [**Fig. 5(b)**]. One also sees that the intensity of the magnon mode, seen in **Fig. 5(b)** as a straight-line trace in between the two electromagnetic modes that avoid crossing, becomes comparable to the intensity of the electromagnetic mode, which is especially seen for $H = 1000..3000$ Oe and $H = 4000..5000$ Oe. Both the stronger anti-crossing effect and the larger magnitude of the magnon mode with respect to the result in **Fig. 5(a)** are attributed to the increased thickness of the YIG layer [**64**].

We also consider the scenario of a standalone hot-pressed polycristalline YIG thick plate having the same dielectric permittivity as the YIG films considered above, as well as the same cross-sectional dimensions as the multilayer in **Fig. 1**. Note that the HDC layers are not needed in this case because the thick YIG plate simultaneously plays the role of the resonator for the magnon mode and the resonator for the electromagnetic mode. It is known that the magnetic damping for the hot-pressed YIG is larger than the value $\alpha_G = 0.001$ for high-quality single crystal YIG film, which has been used in our simulations. The dielectric permittivity may also be different. However, we keep using $\alpha_G = 0.001$ and $\varepsilon = 15$ for the sake of consistency and note that in our model for larger value of $\alpha_G$ one should expect the same result but with larger linewidth and lower magnitude of the magnon mode peak.

The result for the YIG thick plate is shown in **Fig. 5(c)**. Naturally, the anti-crossing effect is very strong because of a larger number of spins is involved in the photon-magnon interaction and due to the fact that all energy of the microwave magnetic field is now concentrated in the magnetic layer.

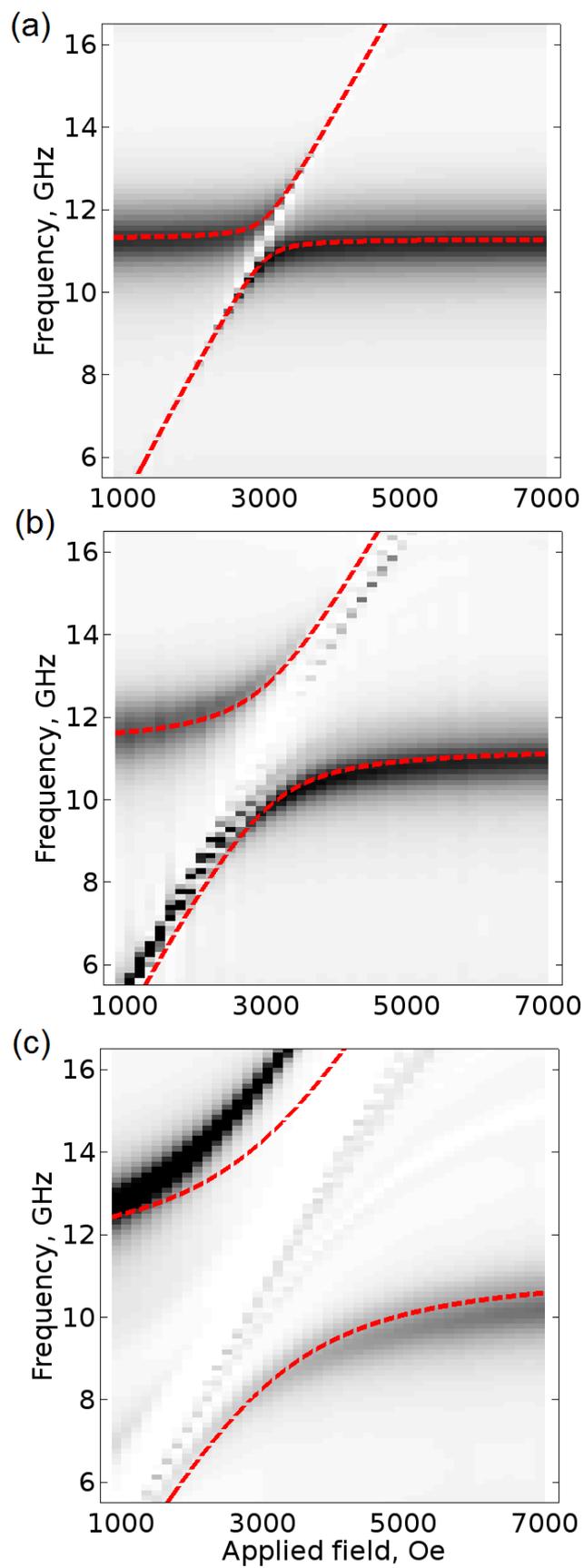

**FIG. 5** Grey-scale maps of simulated spectra of the multilayer as a function of the applied

field and microwave frequency. (a) 50 μm-thick YIG film, (b) 500 μm-thick YIG film, and (c) all-YIG structure. Dashed line denote the best fits obtained using Eq. (3). White colour – zero intensity, black colour – maximum of intensity. The inaccuracy of the fitting of the upper branch of the photon mode in Panel (c) is probably because this branch extends into the frequency region of the higher-order electromagnetic modes of the multilayer.

From the simulated gray-scale maps in **Fig. 5** we extract a value that characterises the strength of coupling of the electromagnetic resonance mode of the multilayer to the magnon mode. In order to compare the degree of co-operativity of different resonance systems, one traditionally uses an abstract model of two coupled resonators (see , e.g, [**23, 24, 35**]). This standard model reads

$$f_{1(2)} = \frac{f_1^0 + f_2^0}{2} \pm \sqrt{\left( \frac{f_1^0 - f_2^0}{2} \right)^2 + \Delta^2} \, , \tag{3}$$

where $f_1$ and $f_2$ are the frequencies of the coupled resonances, $f_1^0$ and $f_2^0$ are the respective resonance frequencies in the absence of coupling and $\Delta$ is the coupling strength. The coupling strength is measured in frequency units. Extracting $\Delta$ from the best fits of simulated data with this abstract model allows one to formally compare systems with different physical origins. Hereafter, we assume that coupling of the multilayer resonance to the microstrip feeding line is broadband and does not depend on the frequency and applied field within the frequency range of interest. This valid assumption allows one to assign the effect seen in **Fig. 5** to the coupling of the photon mode of the multilayer to the magnon mode. We also assume that $f_2^0$ is the frequency of the pure magnon mode given by the Kittel formula Eq. (2). Hence it depends on the applied field $H$. Similarly, we assume that $f_1^0$ is the frequency of the multilayer mode at $H = 0$. Therefore, it is independent of the applied field.

The dashed lines in **Fig. 5(a-c)** are the best fits of the simulated data with Eq. (3). From the fits in **Fig. 5(a)** we obtain $\Delta = 500$ MHz or $\Delta / f_1^0 = 4.5\%$. The former value is slightly larger and the latter value is 1.5 times smaller than those (450 MHz and 6.8%) for an SRR-YIG system investigated is Ref. [**27**]. This comparison is given as a guide only, because the thickness of the YIG film in Ref. [**27**] was two times smaller (25 μm), but the frequency of the anti-crossing was approximately two times lower ($f_1^0 \approx 6$ GHz). For the thicker 500 μm YIG slab we obtain $\Delta = 1500$ MHz or $\Delta / f_1^0 = 13.5\%$. Finally, for a bulk YIG crystal we obtain $\Delta = 3500$ MHz or $\Delta / f_1^0 = 31.25\%$, which is ~1.5 times higher that for the bulk YIG in Ref. [**16**] (~20%).

In overall, the values of $\Delta$ obtained for the YIG layers of different thickness demonstrate the previously observed trend [16]: the usage of thicker layers leads to stronger photon-magnon interaction and thus a wider tuning range. Furthermore, on average the simulated values of the coupling parameter $\Delta$ are the same as in the SRR-based structures. This allows us to draw the conclusion that insulators can be successfully used instead of metals as the constituent material of tuneable electromagnetic metamaterials. Most significantly, such non-metallic metamaterials can outperform their metal-based counterparts because they will be intrinsically low-loss [9, 37, 38]. In addition, the absence of metallic components may help to resolve electromagnetic compatibility problems often encountered in non-metal structures [22].

Finally, the all-dielectric design holds the potential to be frequency scalable, i.e. to some extent an increase in the operating microwave frequency should be possible by reducing the cross-section of the multilayer. Of course, due to relatively low saturation magnetisation of YIG the operation at higher microwave frequencies will require larger values of the applied static magnetic field $H$. The same larger values of $H$ would be required for an SRR-YIG device (or any other magnetic devices containing YIG exploiting the FMR) to operate at higher frequencies. It should be noted that at higher frequencies SRR-based devices may suffer even more from increased losses in the SRR structure. In contrast, this is not expected in our magneto-insulating multilayered structure.

On the other hand, the requirement for the applied magnetic field can be relaxed if one uses spinel or hexaferrite materials [65] instead of YIG. These materials are characterised by a large saturation magnetisation of ~3000–5000 G and their magnetic losses given by $\alpha_G = 0.002$ or so. The dielectric permittivity of spinels and hexaferrites is ~12, which is lower than that of YIG assumed throughout this paper. However, this is not a problem because the thin film is surrounded by two HDC layers. Our simulations show that the magnetic field intensity profile remains unchanged even if the two HDC layers are separated by a 100 µm-high gap. This suggests that one can use insulating thin-film magnetic materials with a relatively low dielectric permittivity without changing the operating frequency and significantly loosing in the magnetic field enhancement.

CONCLUSIONS

Our rigorous numerical simulations have revealed the potential of all-magneto-dielectric multilayered structures to outperform conventional split-ring-resonator-based magnetically tuneable metamaterials and quantum systems. The proposed multilayer structure combines standard and commercially available constituent materials such as YIG and high dielectric constant materials, and it can be driven by both near- and far-field zone microwave magnetic signals produced by different types of sources. The flexibility of the design allows one to employ other magneto-insulating materials instead of YIG, even if the dielectric permittivity of those materials is relatively low as compared with that of the non-magnetic materials. As shown in Ref. [2], decent optical properties of YIG make it possible to create hybrid microwave-optical devices. The presence of non-magnetic dielectric materials additionally improves the optical properties of the investigated multilayer thus making it potentially attractive for the application in, e.g., recently proposed microwave quantum illumination [29].

ACKNOWLEDGMENTS

This work was supported by the Australian Research Council. ISM gratefully acknowledges a postdoctoral research fellowship from the University of Western Australia.

REFERENCES

[1] A. G. Gurevich and G. A. Melkov, *Magnetization oscillations and waves* (CRC, Boca Raton, Florida, 1996).

[2] D. D. Stancil and A. Prabhakar, *Spin Waves: Theory and Applications* (Springer, Berlin, 2009).

[3] V. V. Kruglyak, S. O. Demokritov, and D. Grundler, J. Phys. D: Appl. Phys. **43**, 264001 (2010); doi: 10.1088/0022-3727/43/26/264001

[4] A. A. Serga, A. V. Chumak, and B. Hillebrands, J. Phys. D: Appl. Phys. **43**, 264002 (2010); doi: 10.1088/0022-3727/43/26/264002

[5] T. Ueda and M. Tsutsumi, IEEE Trans. Magnet. **41**, 3532 (2005); doi: 10.1109/INTMAG.2005.1463707

[6] Y. He, P. He, S. D. Yoon, P. Parimi, F. Rachford, V. Harris, and C. Vittoria, J. Magn. Magn. Mater. **313**, 187 (2007); doi:10.1016/j.jmmm.2006.12.031

[7] C. Caloz and T. Itoh, *Electromagnetic Metamaterials: Transmission Line Theory and Microwave Applications* (John Wiley and Sons, New York, 2006).

[8] M. Tsutsumi and T. Ueda, in *Microwave Symposium Digest, 2004 IEEE MTT-S International*, Vol. 1 (2004) pp. 249–252 Vol.1. doi: 10.1109/MWSYM.2004.1335859

[9] Q. Zhao, J. Zhou, F. Zhang, and D. Lippens, Mater. Today **12**, 60 (2009); doi: 10.1016/S1369-7021(09)70318-9

[10] K. Bi, Y. Guo, X. Liu, Q. Zhao, J. Xiao, M. Lei, and J. Zhou, Sci. Rep. **4**, 7001 (2014); doi: 10.1038/srep07001

[11] H. Zhao, J. Zhou, Q. Zhao, B. Li, L. Kang, and Y. Bai, Appl. Phys. Lett. **91**, 131107 (2007); http://dx.doi.org/10.1063/1.2790500

[12] F. Rachford, D. Armstead, V. Harris, and C. Vittoria, Phys. Rev. Lett. **99**, 057202 (2007); http://dx.doi.org/10.1103/PhysRevLett.99.057202

[13] A. Pimenov, A. Loidl, K. Gehrke, V. Moshnyaga, and K. Samwer, Phys. Rev. Lett. **98**, 197401 (2007); http://dx.doi.org/10.1103/PhysRevLett.98.197401

[14] L. Kang, Q. Zhao, H. Zhao, and J. Zhou, Opt. Express **16**, 8825 (2008); http://dx.doi.org/10.1364/OE.16.008825

[15] J. N. Gollub, J. Y. Chin, T. J. Cui, and D. R. Smith, Opt. Express **17**, 2122 (2009); http://dx.doi.org/10.1364/OE.17.002122

[16] G. B. G. Stenning, G. J. Bowden, L. C. Maple, S. A. Gregory, A. Sposito, R. W. Eason, N. I. Zheludev, and P. A. J. de Groot, Opt. Express **21**, 1456 (2013); http://dx.doi.org/10.1364/OE.21.001456

[17] S. A. Gregory, G. B. G. Stenning, G. J. Bowden, N. I. Zheludev, and P. A. J. de Groot, New J. Phys. **16**, 063002 (2014); http://iopscience.iop.org/1367-2630/16/6/063002

[18] Y. Huang, G. Wen, W. Zhu, J. Li, L.-M. Si, and M. Premaratne, Opt. Express **22**, 16408 (2014); http://dx.doi.org/10.1364/OE.22.016408

[19] T. Kurihara, K. Nakamura, K. Yamaguchi, Y. Sekine, Y. Saito, M. Nakajima, K. Oto, H. Watanabe, and T. Suemoto, Phys. Rev. B **90**, 144408 (2014); http://dx.doi.org/10.1103/PhysRevB.90.144408

[20] G. He, Rui-xin Wu, Y. Poo, and P. Chen, J. Appl. Phys. **107**, 093522 (2010); http://dx.doi.org/10.1063/1.3359718

[21] R. X. Wu, J. Appl. Phys. **97**, 076105 (2005); http://dx.doi.org/10.1063/1.1883718

[22] E. Lheurette, *Metamaterials and Wave Control* (Wiley, London, 2013).

[23] Y. Tabuchi, S. Ishino, T. Ishikawa, R. Yamazaki, K. Usami, and Y. Nakamura, Phys. Rev. Lett. **113**, 083603 (2014); http://dx.doi.org/10.1103/PhysRevLett.113.083603

[24] X. Zhang, C.-L. Zou, L. Jiang, and H. X. Tang, Phys. Rev. Lett. **113**, 156401 (2014); http://dx.doi.org/10.1103/PhysRevLett.113.156401

[25] M. Goryachev, W. G. Farr, D. L. Creedon, Y. Fan, M. Kostylev, and M. E. Tobar, Phys. Rev. Applied **2**, 054002 (2014); doi: 10.1103/PhysRevApplied.2.054002

[26] Y. Cao, P. Yan, H. Huebl, S. T. B. Goennenwein, and G. E. W. Bauer, "Magnon-polaritons in microwave cavities," (2014), arXiv:1412.5809 [cond-mat.mes-hall].

[27] B. Bhoi, T. Cliff, I. S. Maksymov, M. Kostylev, R. Aiyar, N. Venkataramani, S. Prasad, and R. L. Stamps, J. Appl. Phys. **116**, 243906 (2014); http://dx.doi.org/10.1063/1.4904857

[28] H. Huebl, C. Zollitsch, J. Lotze, F. Hocke, M. Greifenstein, A. Marx, R. Gross, and S. Goennenwein, Phys. Rev. Lett. **111**, 127003 (2013); http://dx.doi.org/10.1103/PhysRevLett.111.127003

[29] S. Barzanjeh, S. Guha, C. Weedbrook, D. Vitali, J. H. Shapiro, and S. Pirandola, Phys. Rev. Lett. **114**, 080503 (2015), http://dx.doi.org/10.1103/PhysRevLett.114.080503

[30] E. D. Lopaeva, I. Ruo Berchera, I. P. Degiovanni, S. Olivares, G. Brida, and M. Genovese, Phys. Rev. Lett. **110**, 153603 (2013); http://dx.doi.org/10.1103/PhysRevLett.110.153603

[31] D. Smith, W. Padilla, D. Vier, S. Nemat-Nasser, and S. Schultz, Phys. Rev. Lett. **84**, 4184 (2000); http://dx.doi.org/10.1103/PhysRevLett.84.4184

[32] F. Falcone, F. Martín, J. Bonache, R. Marqués, and M. Sorolla, Microwave Opt. Technol. Lett. **40**, 3 (2004); doi: 10.1002/mop.11269

[33] V. Cherepanov, I. Kolokolov, and V. Lvov, Phys. Rep. **229**, 81 (1993); doi: 10.1016/0370-1573(93)90107-O

[34] A. Imamoglu, Phys. Rev. Lett. **102**, 083602 (2009); http://dx.doi.org/10.1103/PhysRevLett.102.083602

[35] S. Probst, N. Kukharchyk, H. Rotzinger, A. Tkalčec, S. Wünsch, A. D. Wieck, M. Siegel, A. V. Ustinov, and P. A. Bushev, Appl. Phys. Lett. **105**, 162404 (2014); http://dx.doi.org/10.1063/1.4898696

[36] For example, one can use commercial Eccostock HIK 500 dielectric materials designed for low-loss GHz microwave operation and having the dielectric permittivity in between 3 and 30 including 15, which corresponds to that of YIG in our model. These materials are available in different shapes and sizes suitable for the application in the

proposed multilayered structure.

[37] S. O'Brien and J. B. Pendry, J. Phys.: Condens. Matter **14**, 4035 (2002); doi: 10.1088/0953-8984/14/15/317

[38] B.-I. Popa and S. A. Cummer, Phys. Rev. Lett. **100**, 207401 (2008), http://dx.doi.org/10.1103/PhysRevLett.100.207401

[39] G. Boudarham, R. Abdeddaim, and N. Bonod, Appl. Phys. Lett. **104**, 021117 (2014), http://dx.doi.org/10.1063/1.4861166

[40] G. Counil, Joo-Von Kim, T. Devolder, C. Chappert, K. Shigeto, and Y. Otani, J. Appl. Phys. **95**, 5646 (2004); http://dx.doi.org/10.1063/1.1697641

[41] I. S. Maksymov and M. Kostylev, Physica E: Low-dimensional Systems and Nanostructures **69**, 253 (2015); http://dx.doi.org/10.1016/j.physe.2014.12.027

[42] V. E. Demidov, B. A. Kalinikos, and P. Edenhofer, J. Appl. Phys. **91**, 10007 (2002); http://dx.doi.org/10.1063/1.1475373

[43] A. B. Ustinov, V. S. Tiberkevich, G. Srinivasan, A. N. Slavin, A. A. Semenov, S. F. Karmanenko, B. A. Kalinikos, J. V. Mantese, and R. Ramer, J. Appl. Phys. **100**, 093905 (2006); http://dx.doi.org/10.1063/1.2372575

[44] A. B. Ustinov, G. Srinivasan, and B. A. Kalinikos, Appl. Phys. Lett. **90**, 031913 (2007); http://dx.doi.org/10.1063/1.2432953

[45] A. B. Ustinov, G. Srinivasan, and Y. K. Fetisov, J. Appl. Phys. **103**, 063901 (2008); http://dx.doi.org/10.1063/1.2841200

[46] Y.-Y. Song, J. Das, P. Krivosik, N. Mo, and C. E. Patton, Appl. Phys. Lett. **94**, 182505 (2009); http://dx.doi.org/10.1063/1.3131042

[47] K. L. Livesey and R. L. Stamps, Phys. Rev. B **81**, 094405 (2010); http://dx.doi.org/10.1103/PhysRevB.81.094405

[48] M. A. Popov, I. V. Zavislyak, and G. Srinivasan, J. Appl. Phys. **110**, 024112 (2011); http://dx.doi.org/10.1063/1.3607873

[49] Y. Zhu, G. Qiu, and C. S. Tsai, J. Appl. Phys. **111**, 07A502 (2012); http://dx.doi.org/10.1063/1.3671779

[50] I. V. Bychkov, D. A. Kuzmin, and V. G. Shavrov, J. Magn. Magn. Mater **329**, 142 (2013); doi: 10.1016/j.jmmm.2012.10.021

[51] A. A. Nikitin, A. B. Ustinov, A. A. Semenov, B. A. Kalinikos, and E. Lähderanta, Appl. Phys. Lett. **104**, 093513 (2014); doi: 10.1063/1.4867985

[52] S. Schäfer, A. V. Chumak, A. A. Serga, G. A. Melkov, and B. Hillebrands, Appl. Phys. Lett. **92**, 162514 (2008); doi: 10.1063/1.2917590

[53] A. A. Serga, A.V. Chumak, A. André, G. A. Melkov, A. N. Slavin, S. O. Demokritov, and B. Hillebrands, PRL **99**, 227202 (2007); http://dx.doi.org/10.1103/PhysRevLett.99.227202

[54] B. Z. Rameshti, Y. Cao, and G. E. W. Bauer, Magnetic spheres in microwave cavities, arXiv:1503.02419 [cond-mat.mes-hall]

[55] I. S. Maksymov and M. Kostylev, J. Appl. Phys. **113**, 043927 (2013); http://dx.doi.org/10.1063/1.4789812

[56] I. S. Maksymov and M. Kostylev, J. Appl. Phys. **116**, 173905 (2014); http://dx.doi.org/10.1063/1.4900999

[57] P. R. Emtage, J. Appl. Phys. **49**, 4475 (1978); http://dx.doi.org/10.1063/1.325452

[58] A. Taflove and S. C. Hagness, *Computational Electrodynamics: The Finite–Difference Time–Domain Method*, 3rd ed. (Artech House Publishers, Boston, 2005).

[59] S. S. Kalarickal, P. Krivosik, M. Wu, C. E. Patton, M. L. Schneider, P. Kabos, T. J. Silva, and J. P. Nibarger, J. Appl. Phys. **99**, 093909 (2006); http://dx.doi.org/10.1063/1.2197087

[60] R. D. McMichael and P. Krivosik, IEEE Trans. Magnet. **40**, 2 (2004); doi: 10.1109/TMAG.2003.821564

[61] K. Yu. Guslienko, S. O. Demokritov, B. Hillebrands, amd A. N. Slavin, Phys. Rev. B **66**, 132402 (2002); http://dx.doi.org/10.1103/PhysRevB.66.132402

[62] V. F. Dmitriev and B. A. Kalinikos, Sov. Phys. J. **31**, 875 (1988); http://dx.doi.org/10.1007/BF00893541

[63] In BFMR measurements one often discusses the complex scattering parameter $S_{21}$ obtainable, for example, from the values of the microwave electric field in the stripline.

However, by virtue of the Maxwell's equations solved consistently with the LLG equation, one obtains the same spectral dependence by using the microwave magnetic field inside the magnetic film.

[64] In our simulations of the 50 μm-thick YIG film we have 20 mesh points across the film thickness. However, we have 200 points for the 500 μm-thick YIG film because we keep the same spatial resolution in our simulations. We verified that this difference does not interfere with our analysis.

[65] V. G. Harris, A. Geiler, Y. Chen, S. D. Yoon, M. Wu, A. Yang, Z. Chen, P. He, P. V. Parimi, X. Zuo, C. E. Patton, M. Abe, O. Acher, and C. Vittoria, J. Magn. Magn. Mater. **321**, 2035 (2009); http://dx.doi.org/10.1016/j.jmmm.2009.01.004.